# On the minimal model of kinetic cooperativity. The case of glucokinase

*Leonid N. Christophorov*

Bogolyubov Institute for Theoretical Physics, NAS Ukraine. 14b Metrologichna Str., 03143 Kyiv, Ukraine

Corresponding author: L.N. Christophorov. E-mail: <lchrist@bitp.kiev.ua>

**Abstract**

The minimal 3-state scheme of kinetic cooperativity of monomeric enzymes is subjected to detailed analysis. The rigorous criteria of positive cooperativity and its sigmoidal version are established in terms of the system parameters (rate constants). It is shown that the cooperativity extent is especially sensitive to the rates and direction of the exchange between conformational states of the free enzyme. However, no necessity of the "kinetic resonance" (or, moreover, its generality claimed recently) for enhancing cooperativity is revealed. Overall, while the minimal 3-state model serves well for qualitative understanding the origin of kinetic cooperativity, it is hardly suitable for quantitative describing reactions of real enzymes, as it is shown with the case of glucokinase.

# I. Introduction

Kinetic cooperativity is one of the specific manifestations of enzyme regulatory properties as deviations from the classical behaviour dictated by the evergreen Michaelis-Menten (MM) model [1-3]. The most characteristic feature of kinetic cooperativity is that it can be peculiar even to monomeric enzymes with an only binding site for only one substrate. This sounds somewhat contradictory to the very name of the phenomenon. Indeed, how is it possible to speak about cooperativity (thought within the classical oligomeric/allosteric models of equilibrium binding [4,5]) in the absence of interacting binding sites? Nevertheless, due to the similarity between the sigmoidal saturation curves of the enzymatic reaction velocities and those of equilibrium ligand binding by oligomeric proteins, this name has been adopted for monomeric enzymes, too [6]. The necessity of the adjective "kinetic" should be particularly noted here, as the effect is possible under non-equilibrium conditions only. It is these conditions that are inherent in the enzyme functioning.

The first theoretical models of kinetic cooperativity appeared in the late sixties [7-11], when taking into consideration the conformational (sub)states of the enzyme reaction states became generally recognised. As summarised in subsequent reviews (see e.g. [6,12,13]), the indispensable condition of kinetic cooperativity is the presence of at least two ($E$ and $E^*$) interconverting conformational states of a free enzyme (splitting of the classical MM scheme into two reaction pathways), with different affinities to the substrate $S$. Sometimes, this condition is called "conformational selection" [14]. The interconversions $E^* \rightleftarrows E$ should be biased in such a way that state $E$ with the *lower* affinity should be *more* stable than state $E^*$. Besides, these conformational transitions should be *sufficiently slow* with respect to the enzyme turnover time. Then the physical reason of *positive* cooperativity (as a transition of the enzyme to a more effective functional regime with concentration [$S$] growing) is that, under faster arrival of substrates, the conformational equilibrium between $E$ and $E^*$ has no time to be completed, and the reaction starts to proceed along the less stable, but of higher affinity, channel. In such a way, the enzyme structure memory shows up. Whitehead termed this as interaction between two subsequent substrates – not through space but "through time" [8].

At this point, the qualitative picture of kinetic cooperativity of monomeric enzymes could be thought completed. Later, generalisations of the Rabin scheme [7] on more complex schemes with a greater number of intermediate stages and, consequently, more complex mathematics appeared (see, e.g., [15]). They did not essentially change the views on the nature of the effect, though. As a real example, glucokinase (GCK) has been cited most often. For this enzyme, the presence and physiological significance of the positive cooperativity were undoubtedly revealed [16-18]. Recently, however, new experimental works on this important enzyme have appeared [19,20]. In them, remaining within the traditional interpretation of the effect (now proposed to be termed "allokairy" [21]), the authors suggest an additional condition of the "kinetic resonance" (namely, coincidence of the values of rates of the catalytic stage and those of the conformational exchange) needed for the optimisation of cooperativity. This entailed the work [22] in which, in an extremely simplified scheme, the mentioned resonance itself (as an important and *general* condition to improve enzyme regulatory abilities) and its quantitative manifestation in the GCK functioning were theoretically substantiated. However, the method for obtaining these results is set out in [22] somewhat unclearly. Also, both the resonance itself and its claimed generality do not fit into the traditional ideas of kinetic cooperativity. That is why in this paper the minimal model used in [22] is subjected to detailed analysis, with a particular emphasis on the existence of the resonance effect and the possibility of quantitative describing GCK cooperativity.

## II. The model

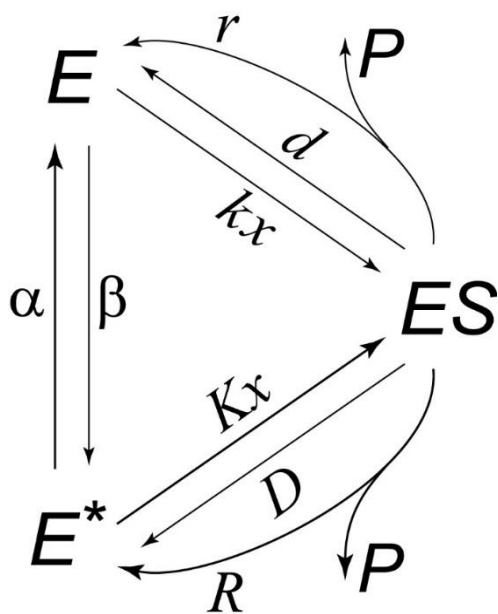

**Fig. 1.** Conformationally splitted Michaelis-Menten scheme analysed in [22]. States $E$, $E^*$ represent those of unliganded GCK that differ in affinities to substrate $S$ (glucose); henceforth, its concentration $[S]$ is denoted as $x$. Here $Kx$, $kx$ are the rate constants of substrate binding, $D$, $d$ are those of unproductive dissociation, $R$, $r$ are those of catalytic conversion of $S$ into product $P$, and $\alpha$, $\beta$ are the rate constants of conformational interconversions.

Presented in Fig. 1, the scheme of converting substrate $S$ into product $P$ is, in fact, a simplified version of Rabin's scheme [7]. It can be considered as a minimal model of kinetic cooperativity for enzymes with a single binding site. Its mandatory element, as noted above, is the presence of two conformational states $E$, $E^*$ of the free enzyme with different affinities to substrate $S$ (henceforth, we assume that $K > k$). While the conformational selection [14] is introduced explicitly, another famous element of protein reactions – induced fit – usually represented by different conformations of complex $ES$ with interconversions like $ES \rightleftarrows E^*S$ is not necessary here (its possible presence can be reflected by the values of corresponding rates). The possibility to illustrate monomeric cooperativity with the help of a triangular scheme like that in Fig. 1 was mentioned earlier (see e.g. [6,23-25]). Its thorough analysis, however, was not performed – perhaps, because of its apparent simplicity. Indeed, the scheme is quite simple but still contains eight parameters. After a tempting attempt of using it for coming to rather fundamental conclusions [22], such an analysis does not seem superfluous at all.

The evolution equations for probabilities $P_E(t), P_{E^*}(t), P_{ES}(t)$ of the corresponding states in Fig. 1 read:

$$\frac{dP_E}{dt} = -(\beta + kx)P_E + \alpha P_{E^*} + (d+r)P_{ES}$$
$$\frac{dP_{ES}}{dt} = kxP_E + KxP_{E^*} - (D+R+d+r)P_{ES}$$
(1)

with the conservation condition $P_E(t) + P_{E^*}(t) + P_{ES}(t) = 1$ for any $t$.[1] Solving them in a trivial way, for the stationary reaction velocity per enzyme molecule $v = (R+r)P_{ES} \equiv k_{cat}P_{ES}$, one has

$$\frac{v}{k_{cat}} = \frac{x^2 + Cx}{x^2 + x(B+C) + AC},$$
(2)

or, as it is presented in [22],

$$\frac{1}{v} = \frac{1}{k_{cat}}\left(1 + \frac{A}{x} + \frac{B-A}{x+C}\right),$$
(3)

where $k_{cat} = R + r$, and

$$A = \frac{(\alpha + \beta)(D + d + R + r)}{\alpha k + \beta K},$$
$$B = \frac{K(d+r) + k(D+R)}{Kk},$$
(4)
$$C = \frac{\alpha k + \beta K}{Kk}.$$

The rate constants $\alpha, \beta, D, R, d, r$ are supposed to be measured in s$^{-1}$. If to measure substrate concentration $x \equiv [S]$ in mM, then $K, k$ are measured in s$^{-1}$mM$^{-1}$. Being interested in cooperativity, that is, in the behaviour of function $v(x)$, one can see from Eqs.(2,3) that it is determined by the three combinations of the model parameters, $A, B, C$, all positive.[2] Note also that quantities $A, B, C$ are invariant to simultaneous multiplication of all the rate constants by the same arbitrary factor.

---

[1] After imposing the conservation condition, the set (1) is identical to the standard chemical kinetics equations for concentrations $[E], [E^*], [ES]$ under the condition $[E] + [E^*] + [ES] = [E_t]$, where $[E_t]$ is the total enzyme concentration [26].

Cooperativity means a specific deviation of $v(x)/k_{cat}$ from the MM hyperbola $x/(x+K_M)$, where $K_M$ is Michaelis' constant. Here it should be noted that Eq.(2) excludes the possibility of substrate inhibition, since the first derivative $v'(x)$ can be zero if only $Bx^2 + 2ACx + AC^2 = 0$. The latter equation, however, has no positive root under $A, B, C$ all positive. That is why the triangular scheme in Fig. 1 can really be viewed as the minimal model distinguishing precisely the monomeric cooperativity phenomenon.

Next, one can easily see that

$$\frac{x}{x+A} < \frac{x^2+Cx}{x^2+x(B+C)+AC} < \frac{x}{x+B} \quad \text{if } A > B$$

$$\frac{x}{x+B} < \frac{x^2+Cx}{x^2+x(B+C)+AC} < \frac{x}{x+A} \quad \text{if } A < B$$

(5)

for *any* $C$ and $x > 0$.

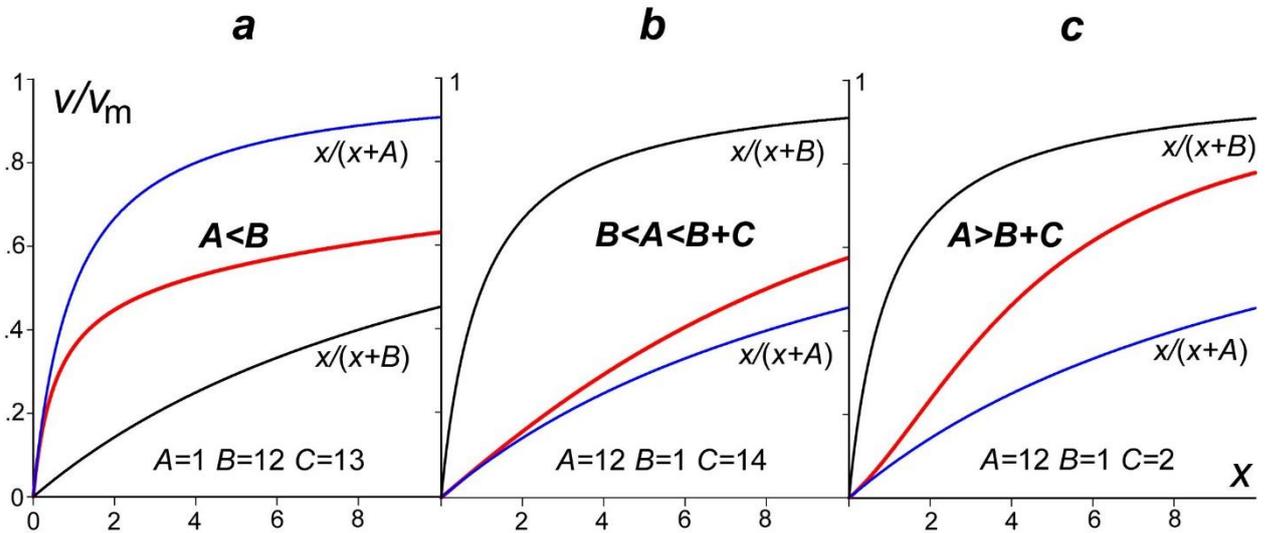

**Fig. 2.** Saturation curves in the cases of: (*a*) negative cooperativity, $A < B$; (*b*) positive non-sigmoidal cooperativity, $A > B$ but $A < B+C$; (*c*) positive sigmoidal cooperativity, $A > B+C$.

---

[2] The case of the absence of conformational transitions, $\alpha = \beta = 0$ (when $A$ becomes indefinite) corresponds to the stationary solution to set (1) with $v(x) \sim x/(x+B)$, i.e., to the MM hyperbola.

Inequalities (5) mean that the curve $v(x)$ is always situated between two MM hyperbolae. If $A > B$, then it moves from the lower curve $x/(x+A)$ (with which it coincides in the limit $x \to 0$) to the upper curve $x/(x+B)$ in the limit $x \to \infty$. On the contrary, if $A < B$, then such a transition proceeds from the upper curve $x/(x+A)$ to the lower one, $x/(x+B)$, see Fig. 2a,b. The first case corresponds to positive cooperativity, whereas the second – to negative cooperativity. The case of positive cooperativity needs further classifying, though. It is desirable to have a quantitative measure of cooperativity. For this, the Hill coefficient is taken most often.

## III. The Hill coefficient. Cooperativity and sigmoidicity

Formally, the Hill coefficient is defined as

$$n_H = d\log\left(\frac{v}{v_m - v}\right) / d\log x, \tag{6}$$

where $v_m = v(x = \infty) = R + r = k_{cat}$ in our case. Cooperativity is positive if $n_H$ exceeds 1; the more $n_H$, the more pronounced positive cooperativity. If $n_H < 1$, they say about negative cooperativity. The Hill coefficient is constant only in the case of Hill's equation that corresponds to the unreal case of one-stage binding of $n$ ligands, $E + nS \to ES_n$, when the saturation curve is described by the formula $\theta(x) = x^n / (x^n + x_{0.5}^n)$. In other cases, our included, $n_H(x)$ depends on concentration $x$. Then they either introduce the effective coefficient obtained by nonlinear regression of $v(x)$ to the form of $\theta(x)$, or use the maximum value $n_{H\max} = \max\{n_H(x)\}$. According to Eq.(6), for $v(x)$ represented by Eq.(2), $n_H(x)$ in terms of $A, B, C$ reads:

$$n_H(x) = \frac{Bx^2 + 2ACx + AC^2}{Bx^2 + (A+B)Cx + AC^2}. \tag{7}$$

It is easy to see that if $A > B$, then $n_H(x) \geq 1$, going to unity in the limits $x \to 0, \infty$. Its maximal value reached at concentration $x_{\max} = C\sqrt{A/B}$ is equal to

$$n_{H\max} = 2\frac{\sqrt{A}}{\sqrt{A} + \sqrt{B}} = 2\frac{\sqrt{A/B}}{1 + \sqrt{A/B}} < 2$$

and is always less than 2 (that reflects the fact that the reaction model includes only two conformational channels).

Thus, from either Eq.(5) or Eq.(7), the same positive cooperativity condition follows: $A > B$ for *any* $C$.[3] However, the quantity $C$ plays an important role, too, and not only for position $x_{max}$ of maximum value $n_{H\,max}$. The fact is that the transition from curve $x/(x+A)$ to curve $x/(x+B)$ can occur in a sigmoidal fashion, as depicted in Fig. 2c. Actually, it is the sigmoidicity that is practically always implied behind positive cooperativity, see e.g. [14].[4] This is understandable, since the initial interpretations of cooperative binding were often given within the Hill equation, and curve $\theta(x) = x^n / (x_{0.5}^n + x^n)$ always has a flection at $x^* = x_{0.5}\left[(n-1)/(n+1)\right]^{1/n}$ if $n > 1$. Besides, the trigger character (more pronounced with $n$ growing) of a sigmoidal curve indicates the possibility of a transition to another binding/reaction regime within a narrower concentration interval, thereby enhancing the regulatory capability.[5] This important subset of saturation curves is characterised by the existence of a positive root to the equation $v''(x) = 0$. According to Eq.(2), this means that

$$Bx^3 + 3ACx^2 + 3AC^2 x + AC^2(B + C - A) = 0. \qquad (8)$$

As $A, B, C > 0$, cubic equation (8) has a positive root if only the last term in its l.h.s. is negative, that is, if

$$A > B + C. \qquad (9)$$

Thus, the condition of sigmoidal cooperativity includes quantity $C$. Then the cooperativity 'phase diagram' looks like shown in Fig. 3.

Now, let us turn to the initial parameters of the model.

## IV. Cooperativity in terms of the system parameters

To begin with, we note that, as it follows from Eq.(4), under equal binding rates, $K = k$, quantities $A$ and $B$ become equal, too: $A = B = (D + R + d + r)/k$. According to Eqs.(2,3,5), this automatically entails the MM dependence $v(x)/v_m = x/(x+A)$ for *any* rates of conformational interconversions $\alpha, \beta$. That is, in accordance with the conventional picture of monomeric cooperativity, we have to consider the cases of different affinities in states $E$ and $E^*$ of the free enzyme. As mentioned earlier, we assume that the binding to state $E^*$ is faster than to state $E$,

---

[3] This condition is cited in [22] as a particular result of a rather complex analysis of a scheme with arbitrary numbers of intermediate states and conformational channels [16]. In the present work, the origin of the cooperativity condition looks more transparent. Besides, in [22] the role of quantity $C$ was paid practically no attention.
[4] Although the possibility of non-sigmoidal positive cooperativity was noted in [12]
[5] just like haemoglobin regulates the binding/release of oxygen under a relatively small partial pressure difference in venous and arterial vessels

i.e. $K > k$. The condition of *positive* cooperativity (which we are predominantly interested in), $A > B$, as it follows from Eq.(4), reads:

$$\alpha k(D+R) > \beta K(d+r). \tag{10}$$

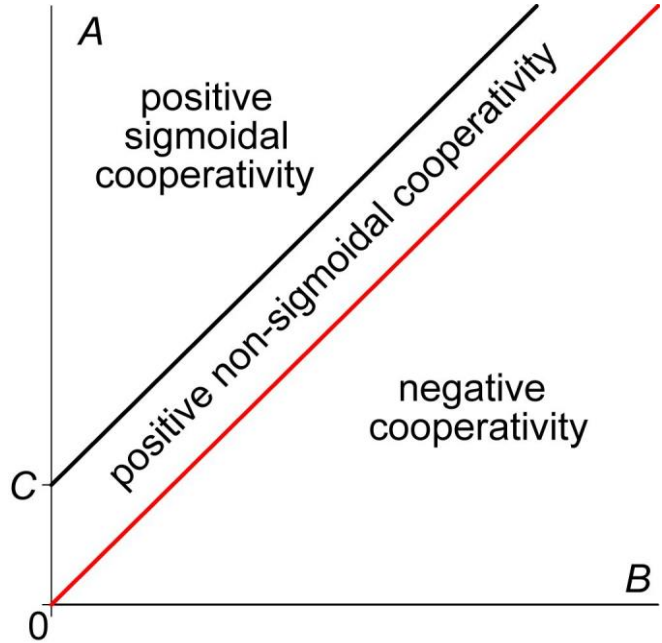

**Fig. 3.** Phase diagram of cooperativity. The tilt angle of the straight lines is $\pi/4$.

Eq.(10) means, in particular, the violation of detailed balance in the triangular scheme in Fig. 1 – as it should be for kinetic cooperativity to exist [6,12,14,15,22]. As for sigmoidicity, its condition (9) acquires a more complex form:

$$(K-k)\left[\alpha k(D+R) - \beta K(d+r)\right] > (\alpha k + \beta K)^2. \tag{11}$$

It is worth to consider some particular cases of these inequalities. Recall that the traditional interpretation of monomeric positive cooperativity implies that state $E$ of lower affinity is more stable. This means that $\alpha$ should be greater than $\beta$ in interconversions $E^* \underset{\beta}{\overset{\alpha}{\rightleftarrows}} E$. At the same time, these interconversions should be sufficiently slow, that is, $\alpha$ should be bounded from top. It is the conditions that allow one to understand the physics of "allokairy". Below we will see that these qualitative considerations are mainly valid, although not absolutely.

In the limiting cases of unidirectional conformational relaxation[6] it is obvious from condition (10) that if $\alpha = 0$ but $\beta > 0$, then the positive cooperativity is impossible. On the

---
[6] Such a unidirectional conformational relaxation could seem nonphysical. However, similar irreversible stages often take place in models of enzymatic reactions. Under violations of detailed balance, they are definitely acceptable. Practically, this is simply a statement that

contrary, if $\beta = 0$ but $\alpha > 0$, then it takes place for *any* $\alpha$. Consider the particular cases in more detail.

(i) $\underline{\beta = 0.}$ Then $A = (D + R + d + r)/k$, $B = [k(D+R) + K(d+r)]/Kk$, and $C = \alpha/K$. The condition $A > B$ holds if $K > k$ – that is, always, see Eq.(10). Note that here $\alpha$ can be arbitrarily large; nevertheless, the positive cooperativity (with the Hill coefficient notably greater than 1) takes place, as $A$ and $B$ do not depend on $\alpha$. This intuitively contradicts the aforementioned traditional requirement of 'slow' conformational transitions and hints once more that behind the positive cooperativity they most often imply its sigmoidal version. For the latter, as it follows from inequality (11) simplified here to the form

$$\alpha < \left(\frac{K}{k} - 1\right)(D + R), \tag{12}$$

the needed upper boundary appears, restricting the conformational transition rate. Now this is consistent with the traditional ideas. That is why the scheme in Fig. 1 with unidirectional relaxation ($\beta = 0$), or even with $d = r = 0$ also, can be viewed as the most minimal model exhibiting 'mnemonic'/kinetic cooperativity.

Of course, this does not mean that in the presence of backward relaxation $E \to E^*$ the positive cooperativity is completely impossible. Consider the case of equal relaxation rate constants, $\alpha = \beta$ (for the case $\alpha > \beta$, see Section V).

(ii) $\underline{\alpha = \beta.}$ Then $A = 2(D + R + d + r)/(K + k)$, $B = [k(D+R) + K(d+r)]/Kk$, and $C = \alpha(K + k)/Kk$. Similarly to case (i), $A$ and $B$ do not depend on $\alpha$, and $C$ is proportional to $\alpha$ but becomes greater (thereby diminishing the sigmoidicity area, see Fig. 3).

The positive cooperativity condition $A > B$ holds, if $k(D+R) > K(d+r)$, that is, if $\frac{D+R}{d+r} > \frac{K}{k}$, see also Eq.(10). Since $K > k$, this means that under the same thermodynamic stability of states $E$ and $E^*$, the dissociation (caused by partial catalytic rate $R$ and/or unproductive dissociation rate $D$) of the enzyme-substrate complex to state $E^*$ should be significantly faster than to state $E$. Again, similarly to case (i), the positive cooperativity is possible at any, even arbitrarily large $\alpha$, but the sigmoidal one – within the interval restricted by even a stronger than (12) inequality (13):

---

transition $E^* \to E$ prevails over the reverse one so strongly that the latter can be neglected. For example, in the illustrative schemes of positive cooperativity [11], the corresponding rates differ by six orders of magnitude.

$$\alpha < \frac{K-k}{(K+k)^2}\left[k(D+R)-K(d+r)\right], \qquad \alpha = \beta. \tag{13}$$

Is the positive cooperativity (and, moreover, sigmoidicity) possible, if $\alpha < \beta$?

(iii) $\underline{\alpha < \beta}$. Despite apparent contradiction with the traditional ideas, this is permissible. As it follows from inequality (10), the positive cooperativity is possible, if

$$\frac{D+R}{d+r} > \frac{\beta K}{\alpha k}, \text{ or } \beta < \alpha \frac{k}{K}\frac{D+R}{d+r}. \tag{14}$$

From Eq.(14) it follows that the dissociation of complex *ES* to state $E^*$ should be even faster, than in the previous case (ii). On the other hand, ratio $\beta/\alpha$ should remain within its limits, $1 < \frac{\beta}{\alpha} < \frac{k}{K}\frac{D+R}{d+r}$. To analyse the sigmoidicity condition is a bit harder. Consider inequality (11) with respect to $\alpha$, holding $\beta$ fixed. Then Eq.(11) turns into quadratic inequality

$$\alpha^2 \frac{k^2}{K-k} + \alpha k\left[2\beta K - (K-k)(D+R)\right] + \beta K\left(\frac{\beta K}{K-k} + d + r\right) < 0, \tag{15}$$

which holds if $\alpha_1 < \alpha < \alpha_2$, where $\alpha_{1,2}$ are the roots of the quadratic trinomial in Eq.(15). Obviously, $\alpha_{1,2}$ can be positive if only $2\beta K - (K-k)(D+R) < 0$, or $\beta < \beta' = \frac{(K-k)(D+R)}{2K}$.

On the other hand, roots $\alpha_{1,2}$

$$\alpha_{1,2} = \frac{1}{2k}\left\{(K-k)(D+R) - 2\beta K \pm \sqrt{(K-k)\left[(K-k)(D+R)^2 - 4\beta K(D+R+d+r)\right]}\right\} \tag{16}$$

can be positive only if $\beta < \beta'' = (K-k)(D+R)^2/4K(D+R+d+r)$. Since $\beta'/\beta'' = 2(D+R+d+r)/(D+R)$, then $\beta'' < \beta'$, so that $\beta$ has its upper boundary

$$\beta < \beta'' = \frac{(K-k)(D+R)^2}{4K(D+R+d+r)}. \tag{17}$$

In turn, $\alpha$ has its lower boundary, $\alpha > \alpha_1$. Therefore,

$$\alpha_1 < \alpha < \beta < \beta''. \tag{18}$$

Thus, in the case $\alpha < \beta$ the sigmoidicity takes place only if inequalities (18) are satisfied. Now let us try to estimate the cooperativity extent by the Hill coefficient.

*In case* (i), when $\beta = 0$,

$$\frac{A}{B} = \frac{K(D+R+d+r)}{k(D+R)+K(d+r)} = 1 + \frac{(K-k)(D+R)}{k(D+R)+K(d+r)}.$$

This ratio grows (so does the maximum value of the Hill coefficient $n_{H\max} = 2\sqrt{A/B}/(1+\sqrt{A/B})$) with either $K$ or $(D+R)$ growing. That is, acceleration of the elementary acts along the lower branch of the scheme in Fig. 1 enhances positive cooperativity, irrespective of the value of conformational relaxation rate constant $\alpha$. However, the value of $C = \alpha/K$ (and thereby the position of maximum, $x_{\max} = C\sqrt{A/B}$) can become rather large, so that sigmoidicity condition (9) can be violated, and $x_{\max}$ can fall out the experiment concentration interval. Then, in particular, an attempt to determine the Hill coefficient by nonlinear regression basing on the available part of the saturation curve with concentrations noticeably less than $x_{\max}$ would lead to a certainly understated value, down to a change of the cooperativity sign.[7] Note again, though, that as long as we are interested in cooperativity only, the Hill coefficient in case (i) can be however close to 2 even for $\alpha \to \infty$ – what intuitively contradicts the traditional ideas on the kinetic cooperativity nature.

The variant of case (i) with sigmoidicity is less exotic, as $\alpha$ is restricted by inequality (12). Fig. 4, *left* (curve a) represents a typical experimental saturation curve of human GCK [20] that was used in work [22].

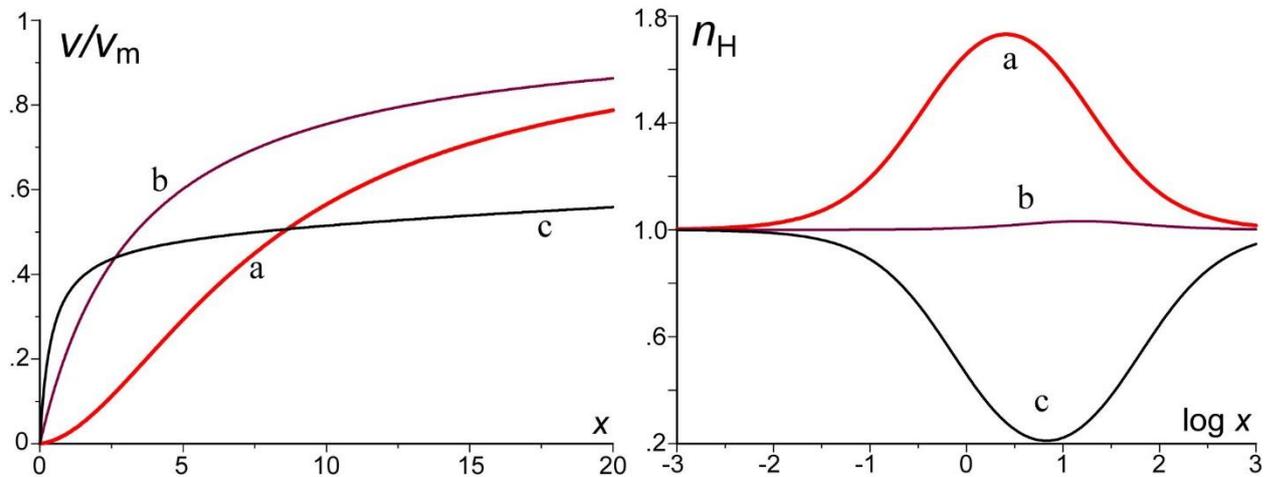

**Fig. 4.** *Left*: Dependence of the reaction velocity on substrate concentration. (a) The saturation curve of human GCK [20]. It can be characterised with the values $A = 125$, $B = 3$, $C = 0.4$ which, in turn, can be ensured by the following set of parameters: $K = 200$, $k = 0.7$, $D = 0.7$, $d = 1$, $R = 85$, $r = 0.8$, $\alpha = 80$, $\beta = 0$; (b-c) the same parameter values except $\alpha = 70$, $\beta = 10$ (curve b), or $\alpha = \beta = R = r = 40$ (curve c). *Right*: The corresponding dependence of the Hill coefficient.

---

[7] not to mention the fact that an attempt to fit a non-sigmoidal curve by nonlinear regression to a sigmoidal one with $n_H > 1$ is hardly consistent

Nonlinear regression to the form of Eq.(2) gives the following values for $A, B, C$: $A = 125$, $B = 3$, $C = 0.4$ which can be ensured by, say, the following set of parameters: $K = 200$, $k = 0.7$, $D = 0.7$, $d = 1$, $R = 85$, $r = 0.8$, $\alpha = 80$, $\beta = 0$ (but see the next Section). Then the peak $n_{H\,\text{max}} \simeq 1.73$ falls into the experimental concentration interval $0 < x < 20$; precisely, $x_{\text{max}} \simeq 2.58$. Also, the experimental curve can be perfectly approximated by the Hill equation $x^{n_H}/(x^{n_H} + x_{0.5}^{n_H})$ with $n_H \simeq 1.61$ and $x_{0.5} \simeq 8.6$. Here, it is important to stress that introducing even a relatively weak backward conformational transition ($\beta \neq 0$) but keeping the interconversion rate constant $k_{\text{ex}} = \alpha + \beta = 80$ unchanged (for example, $\alpha = 70$, $\beta = 10$ instead of $\alpha = 80$, $\beta = 0$) can completely destroy sigmoidicity and sharply diminish $n_H$ (curves *b*). The values $\alpha = \beta = 40$ kill the positive cooperativity at all, converting it into negative, as $A$ becomes less than $B$ (curves *c*).

*In case* (ii) ($\alpha = \beta$), quantities $A$ and $B$ do not depend on $\alpha$ as well, but the conditions become stricter for both positive cooperativity (instead of default condition $K > k$, restriction $1 < K/k < (D+R)/(d+r)$ appears) and sigmoidicity (since the upper boundary for $\alpha$, Eq.(13), is noticeably lower than that in the previous case, Eq.(13)). Besides, $C$ becomes greater, and, consequently, $x_{\text{max}}$, too. Trying to keep the latter not too large, one would obtain typical $n_{H\,\text{max}}$ values like $1.1 \div 1.15$ or smaller. With such $n_{H\,\text{max}}$, sigmoidicity and cooperativity are barely pronounced.

Lastly, *case* (iii) ($\beta > \alpha$) is of mostly academic interest, as state $E^*$ of higher affinity becomes more stable. Under such a condition that, again, is at variance with the traditional ideas of monomeric positive cooperativity, the latter and, especially, sigmoidicity are restricted by rather specific inequalities (16-18). In this case, the Hill coefficient exceeds unity in the second or third decimal places only.

Now we can summarise preliminary conclusions from the analysis of the minimal scheme of kinetic cooperativity, presented by Fig. 1 and Eq.(1).

First, as is obvious from all the formulae above, the catalytic rate constants $R$ and $r$ enter the cooperativity conditions in combinations $(D+R)$ and $(d+r)$ only. This means that $D$ and $R$ (as well as $d$ and $r$) – that is, the rates of unproductive and productive dissociations of the enzyme-substrate complex in each of the reaction channels – play an equal role in the emergence of cooperativity. Although it is often assumed that $D \ll R$ and $d \ll r$, the saturation curve $v(x)/k_{\text{cat}}$ (quantities $A, B, C$) is invariant to permutations $D \rightleftarrows R$ and/or $d \rightleftarrows r$. However, such permutations can radically change $k_{cat} = R + r$, calling into question the necessity of the kinetic resonance $k_{cat} \approx k_{ex}$ (i.e. $\alpha + \beta \approx R + r$) claimed in [20,22] for the optimisation of cooperativity, see also Section V.

Next, the difference between non-sigmoidal and sigmoidal positive cooperativity can be rather pronounced, as is clearly seen in case (i). In it, the positive cooperativity can be quite distinct (of a high Hill coefficient close to 2) due to a large value of the ratio $A/B \gg 1$, but sigmoidicity can be completely absent under too fast conformational relaxation ($\alpha$ exceeds the threshold dictated by Eq.(12)).

In principle, the positive, or even sigmoidal, cooperativity is possible under comparable values of conformation relaxation rates ($\alpha \simeq \beta$), or even, under some restrictions, if $\beta > \alpha$. In these cases, however, it is pronounced rather poorly, or sigmoidicity is absent at all. Overall, one can conclude that the traditional ideas formulated in the 60s-80s on kinetic cooperativity (often identified with sigmoidicity) seem quite reasonable, except in some marginal cases. Namely, the higher-affinity conformational state of the free enzyme should be less stable, and conformational relaxation should not be too fast. While the cooperativity conditions are very sensitive to the ratio $\alpha/\beta$, nowhere in the analysis above a competition of sums $(\alpha + \beta) = k_{ex}$ and $(R + r) = k_{cat}$ shows up. This is important for analysing recent results of the description of GCK cooperativity [20,22], see the next Section.

**V. Kinetic cooperativity of human glucokinase**

Human glucokinase is an enzyme of extraordinary physiological importance since it regulates glucose metabolism. At the same time, it represents the main example of monomeric positive cooperativity. Moreover, the presence of the latter is critical to the organism. If, for some reasons, glucokinase loses such a regulatory property, then it leads to dangerous diabetic-type diseases [17,18,20].

Investigations into GCK cooperativity have a long history. The latest results belong, in particular, to Miller's group [19,20]. In work [20] an experimental sigmoidal saturation curve $v(x)$ was presented, and a 5-state scheme was proposed for quantitative description (although in not too much detail). Basing on the latter, it was stated that the cooperativity is most distinct when the value of catalytic rate $k_{cat}$ is comparable to that of conformational exchange, $k_{ex}$. In the recent theoretical work [22], the authors have made the next interesting step, substantiating the results of work [20] within the triangular scheme pictured in Fig. 1 and claiming the general character of that "kinetic resonance", $k_{cat} \approx k_{ex}$. Let us consider these issues in the light of the results presented above.

Begin with the saturation curve (reproduced in Fig. 4, *left*, curve *a*). It is obviously sigmoidal. As mentioned above, nonlinear regression to the dependence (2) gives the well-defined values $A=125$, $B=3$, $C=0.4$, which of course satisfy the sigmoidicity condition (8). With these values, $n_{H\max} \simeq 1.73$ at $x_{\max} \simeq 2.58$. Regression to the Hill equation (that is, to $x^n/(x^n+a^n)$) gives $n \approx 1.6$ and $a = x_{0.5\text{glucose}} \approx 8.6$. This tells us little about the system rates, however.[8] Turning to Eq.(4), we see that the three equations with the known $A, B, C$ contain eight unknown parameters: $K, k, \alpha, \beta, D, R, d, r$. True, as mentioned above, the last four enter the equations in combination $D+R \equiv G$ and $d+r \equiv g$ only, so that the number of unknowns can be reduced to six. This is still more than one can find from three equations. Nevertheless, having the additional inequalities deduced above, we can try to come to meaningful conclusions concerning the unknowns.

Thus, we start from three equations for unknown $K, k, \alpha, \beta, G, g$ with known $A, B, C$:

$$\begin{cases} \dfrac{(\alpha+\beta)(G+g)}{\alpha k + \beta K} = A \\ \dfrac{Kg + kG}{Kk} = B \\ \dfrac{\alpha k + \beta K}{Kk} = C \end{cases} \quad . \tag{19}$$

To illustrate the way of further analysis, let us consider an even more simplified scheme of monomeric cooperativity, with $\beta = 0$ (case (i)), reducing the number of unknowns to five. Then Eqs.(19) take the form

$$G + g = Ak; \qquad Kg + kG = BKk; \qquad \alpha = CK, \tag{20}$$

and, apart from $K > k$ by default, we have the sigmoidicity condition (12):

$$\alpha < \left(\dfrac{K}{k} - 1\right) G. \tag{21}$$

Solve set (20) with respect to $K, G$ and $g$:

$$K = \alpha/C; \qquad G = \dfrac{(A-B)\alpha k}{\alpha - Ck}; \qquad g = \dfrac{k(\alpha B - ACk)}{\alpha - Ck}. \tag{22}$$

The requirement $G > 0$ is satisfied automatically, as $A > B$ and $\alpha > Ck$ (what is an immediate consequence of inequality $K > k$), whereas the requirement $g > 0$ imposes, as it is easy to check, a stronger restriction on $k$:

---

[8] Besides, it is unclear why the cooperativity ensured by these values is supposed optimal [22] while it is merely an experimental fact, and one can imagine cooperativity with the Hill coefficient even closer to 2 than 1.6.

$$k < \frac{\alpha}{C} \cdot \frac{B}{A}. \tag{23}$$

Now let us see how to satisfy Eqs.(22) and inequality (21) by varying constants $\alpha$ and $k$. Choose arbitrarily $\alpha = 80$.[9] Then it immediately follows from Eq.(22) that $K = 200$, and from Eq.(23) – that $k < 4.8$. Try first a small $k$; for example, let $k$ be equal to 0.7. Then from Eq.(22) one has $G = 85.7$ (and condition (21) is satisfied), and $g = 1.8$ (just these values correspond to those in the Fig. 3 caption). It may seem that the resonance condition approximately holds, since $k_{ex} = (\alpha + \beta) = 80$ and $k_{cat} = (R + r) = 85.8$.[10] However, take $k$ closer to its upper limit, say, $k = 4$. Then $G \simeq 498$ and $g \simeq 2.04$. Note that the cooperativity extent (sigmoidicity, the Hill coefficient, and actually the very curve $a$ in Fig. 3) remains unchanged, but there is no sense to speak about the resonance any longer.

Now proceed to a more realistic case, $0 < \beta < \alpha$, which is indicated by experimental data on GCK (in works [19,20] the following rates of conformational exchange were reported: $\alpha = 435 \text{ s}^{-1}, \beta = 84 \text{ s}^{-1}$). Instead of set (20), we have a full set (19) rewritten as

$$\begin{cases} (\alpha + \beta)(G + g) = ACKk \\ Kg + kG = BKk \\ \alpha k + \beta K = CKk \end{cases} \tag{24}$$

Solve it with respect to $k, G, g$:

$$k = \frac{\beta K}{CK - \alpha}; \qquad g = \beta K \frac{B - \frac{ACK\beta}{(CK - \alpha)(\alpha + \beta)}}{CK - (\alpha + \beta)}; \qquad G = BK - g\frac{CK - \alpha}{\beta}. \tag{25}$$

From the first of these equations it follows that the default condition $K > k$ leads to a lower threshold of $K$, precisely, $K > (\alpha + \beta)/C$, or $CK > (\alpha + \beta)$. Under this restriction and the values $A = 125, B = 3, C = 0.4$, $\alpha = 435, \beta = 84$, it turns out to be impossible to simultaneously satisfy the requirements $G > 0$ and $g > 0$ for any $K > (435 + 84)/0.4 = 1297.5$ (this can be visualised by plotting functions $G(K)$ and $g(K)$ in accordance with Eq.(25), see Fig. 5, *left*). This is not surprising, since the positive cooperativity in the scheme in Fig. 1 is very sensitive to the ratio $\alpha/\beta$ and quickly disappears with $\beta$ growing (see the examples in Fig. 4). Under sufficiently small $\beta \ll \alpha$, set (24) with positive $G, g$ and $k$ still can be satisfied (see

---

[9] This initial choice is of no special importance because, as can be seen from Eqs.(4,19), their numerators and denominators are bilinear in rate constants, so that all the latter can be multiplied by one and the same factor without changing $A, B, C$.

[10] assuming here $R \gg D$ that, strictly speaking, is not obligatory

Fig. 5, *right*) but only if $K \gg k$ and $G \gg g$. For example, keeping $\alpha = 435$ but diminishing $\beta$ to $\beta = 4$, for $K = 2000$ one has $k = 21.9$, $G = 4982$, and $g = 11.6$. This, of course, is well beyond the kinetic resonance condition. What is worse is that one could hardly describe the GCK cooperativity quantitatively with the triangular scheme in Fig. 1 and experimentally measured rate constants $\alpha$ and $\beta$.

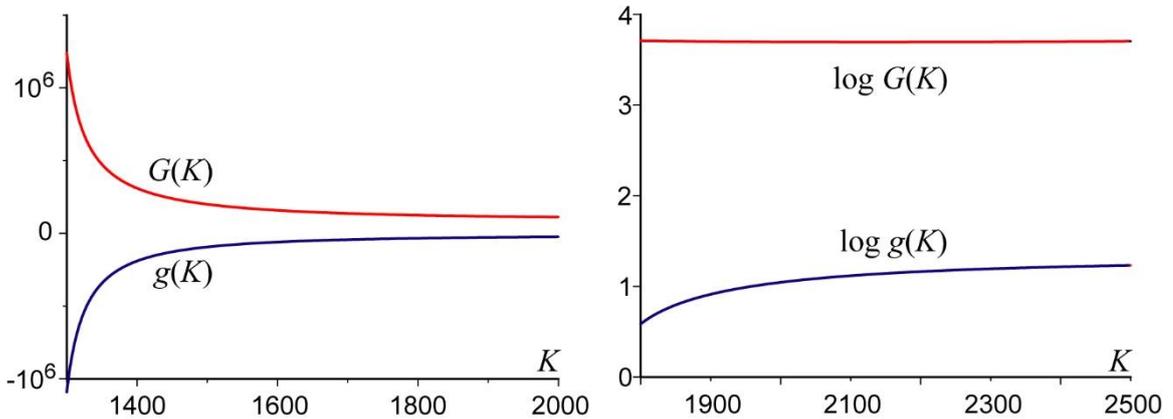

**Fig. 5.** Functions $G(K)$ and $g(K)$ according to Eq.(25). $A = 125, B = 3, C = 0.4$. *Left*: $\alpha = 435, \beta = 84$ [19,20]. *Right*: $\alpha = 435, \beta = 4$.

**VI. Concluding remarks**

Given the fundamental importance of the minimal scheme of kinetic cooperativity, it is subjected to detailed analysis. The rigorous criteria of positive cooperativity and its sigmoidal version are established in terms of the model parameters (rate constants). It is shown, in particular, that the cooperativity extent is very sensitive to the rates and direction of conformational relaxation (in accordance with the traditional qualitative interpretations of kinetic cooperativity of monomeric enzymes). At the same time, no necessity of the kinetic resonance for enhancing the cooperativity extent is revealed. Overall, while the minimal three-state model serves well for qualitative understanding the origin of kinetic cooperativity, it is hardly suitable for quantitative describing the reactions of a concrete real enzyme, as it can be seen in the case of glucokinase. On the other hand, the presented detailed analysis of this minimal model can indicate the ways of modifying structural and kinetic parameters of proteins in order to initiate regulatory properties of the latter [20,27]. Finally, it is worth to note that such properties can be noticeably more pronounced (in particular, in terms of the Hill coefficients noticeably higher than two) in the minimal models of molecular self-organisation with a continuous structural variable, which exploit exactly structural memory of proteins, see e.g. [28,29] and references therein.

**Acknowledgement**

The work is performed within Project 0116U002067 of the NAS of Ukraine.